\documentclass[manuscript]{aastex}
\usepackage{graphics}
\usepackage{graphicx}
\begin{document}
\title{Constraining Stellar Coronal Mass Ejections Through Multi-Wavelength Analysis of the Active M Dwarf EQ Peg}

\author{M. K. Crosley}
\affil{Johns Hopkins University, Department of Physics \& Astronomy, 3400 N. Charles Street, Baltimore, MD 21218}
\and
\author{R. A. Osten}
\affil{Space Telescope Science Institute, 3700 San Martin Dr, Baltimore, MD 21218}
\affil{Center for Astrophysical Sciences, Johns Hopkins University, Baltimore, MD 21218}

\begin{abstract}

Stellar coronal mass ejections remain experimentally unconstrained, unlike their stellar flare counterparts which are observed ubiquitously across the electromagnetic spectrum.
Low frequency radio bursts in the form of a type II burst offer the best means of identifying and constraining the rate and properties of stellar CMEs. 
CME properties can be further improved through the use of proposed solar-stellar scaling relations and multi-wavelength observations of CMEs through the use of type II bursts and the associated flare expected to occur occur alongside. 
We report on 20 hours of observation of the nearby, magnetically active, and well characterized M dwarf star EQ Peg. 
The observations are simultaneously observed with the Jansky Very Large Array at their P-band (230-470 MHz) and at the Apache Point observatory in the SDSS u' filter ($\lambda$ = 3557 \AA).  
Dynamic spectra of the P band data, constructed to search for signals in the frequency-time domains, did not reveal evidence for drifting radio bursts that could be ascribed to type II bursts.  
Given the sensitivity of our observations, we are able to place limits on the brightness temperature and source size of any bursts which may have occurred. 
Using solar scaling rations on four observed stellar flares, we predict CME parameters.
Given the constraints on coronal density and photospheric field strength, our models suggest that the observed flares would have been insufficient to produce detectable type II bursts at our observed frequencies.
We consider the implications of these results, and other recent findings, on stellar mass loss.

\end{abstract}


\section{Introduction}  

M dwarfs are of high interest in the search for exoplanets and the conditions that determine the potential of exoplanet habitability are very diverse and still poorly understood. 
M dwarfs are the most common type of star \citep{Henry2006} and are now known to commonly host exoplanets \citep{Johnson2007,Dressing2013}.
The habitable zone for exoplanets around an M dwarf is $\sim10$ times closer than the habitable zone of the Sun due to its cooler temperature.
The close proximity makes exoplanets more susceptible to magnetic eruptive events and potential space weather impacts.  
Exoplanets residing in the habitable zone of an M dwarf are expected to be tidally locked \citep{Kite2011} which results in a weakened protective magnetosphere and increases their susceptibility to space weather and its deleterious affects \citep{Lammer2007,Khodachenko2007}. 
CMEs are the largest contributor to space weather in the Earth-Sun system \citep{Badruddin2016}.  
CMEs have the potential to erode the atmosphere of exoplanets and prevent habitability \citep{Khodachenko2007}.
Active M dwarfs are known to have a high flaring rate \citep{Lacy1976} and solar relationships predict that these flares are accompanied by a high rate of coronal mass ejections (CMEs) \citep{Yashiro2006}.

Stellar flares are routinely observed on cool stars but clear signatures of stellar CME's have been less forthcoming \citep{Osten2017}.
Solar coronagraphs routinely observe solar CMEs propagating into interplanetary space; current astronomical coronagraphs cannot achieve sufficient star contrast to detect a CME \citep{Mawet2012}, making it an infeasible method for stellar CME observations.  
The type II radio burst is currently the best way of detecting \citep{Osten2017} and constraining CME properties such 
as rate of occurrence and velocity.
CMEs traveling through the corona at sufficiently high speeds, speeds faster than the Alfv\'{e}n speed, drive a magnetohydrodynamic (MHD) shock which produces the emission of a type II burst \citep{Gopalswamy2006}.  
The shocks accelerate non-thermal electrons, which in turn produce radio emission at the fundamental and harmonic of the local plasma frequency via well-known plasma processes \citep{Gopalswamy2006}. 
They appear as a slowly drifting radio burst following an exponential-like path in frequency and time related to the speed of the shock (and thus the speed of the CME) and the ambient density of the local corona \citep{Gopalswamy2006}. 
  
Stars have had a variety of different types of variable radio emission at these frequencies.
The type II's distinct shape makes it easily identifiable and a useful diagnostic for monitoring stellar CME's.
So while not every solar CME will produce a type II burst \citep{Yashiro2006}, the fastest CME's do and every type II burst is uniquely generated by a CME.  
\citet{Crosley2016} proposed routine monitoring of type II bursts as part of a long term strategy to gather events and perform a multi-wavelength analysis to best interpret CMEs and constrain their statistical properties. 

We report on 20 hours of simultaneous observations of the active M dwarf EQ Peg.
The Jansky Very Large Array (JVLA) was used for radio observations to detect type II bursts while the Apache Point Observatory (APO) 0.5m optical telescope was used to observe for stellar flares. 
This is an extension of the 12 hours of YZ CMi observations  performed at the Low Frequency Array and reported in \citet{Crosley2016}.

Section 2 discusses the stellar target, EQ Peg, and its properties.
Section 3 discusses the JVLA P-band observations while section 4 discusses the APO optical observations. 
Section 5 describes the results of those observations. 
Section 6 reviews the methodology for interpreting the multi-wavelength observations.  
Section 7 discusses the results of the analysis.
Section 8 interprets the results.
Section 9 concludes.

	
\section{EQ Peg}

EQ Peg (BD+19 5116 , GJ 896) is a nearby active M dwarf flare star binary with an age of 950 Myr \citep{Parsamyan1995}, at a distance of 6.25 pc, and a separation of 5.8 arcsec \citep{Liefke2008}.  
EQ Peg has previously been observed to have radio bursts \citep{Pallavicini1985,Gagne1998}; its flare rate is known \citep{Lacy1976} and its photospheric magnetic field \citep{Morin2008} and coronal properties \citep{Liefke2008} are known.
These qualities make it a fitting candidate to perform observations and provide a means for extrapolating information from type II bursts.  

The system has a flare rate of 1 flare per hour above a U-band flare energy of 5$\times$10$^{31}$ erg \citep{Lacy1976}. 
The combined system has a spectral type of M3.5 with $U$, $V$, and $B$ magnitudes of 12.737, 11.749, and 10.165 respectively \citep{Koen2010}.
EQ Peg A is a M3.5 star with a mass of 0.39 M$_{\odot}$, a radius of 0.35 R$_{\odot}$, a rotation period of 1.061 days and an average Zeeman Doppler Imageing (ZDI) coronal magnetic field strength of 0.48 kG \citep{Morin2008}. 
Similarly, EQ Peg B is a M4.5 star with a mass of 0.25 M$_{\odot}$, a radius of 0.25 R$_{\odot}$, a rotation period of 0.404 days and an average ZDI coronal magnetic field strength of 0.45 kG \citep{Morin2008}. 
\citet{Liefke2008} investigated the coronal properties of EQ Peg; their findings using \ion{O}{7} and \ion{Ne}{9} claim an averaged density of $10.23^{+7.35}_{-4.41} \times 10^{10}$ cm$^{-3}$ and $<17.58 \times 10^{10}$ cm$^{-3}$ for EQ Peg A and B.  
The differential emission measure for each star peaks at 3MK \citep{Liefke2008}.  
The coronal scale height is
\begin{equation}
H = \frac{k_{B}T}{\mu m_{p}g} = 5.01\left(\frac{M_{\star}}{M_\odot}\right)\left(\frac{R_{\star}}{R_{\odot}}\right)^{2} T\times 10^{3} \textrm{ cm/K} 
\label{equation:scaleheight}
\end{equation}	
where $k_{B}$ is the Boltzmann constant, $m_{p}$ is the proton mass, $\mu$ is the mean molecular weight of the particles equal to 0.6, and g is the local gravity of the star which depends on its mass and radius.  
The peak differential emission measure of T = 3 MK predicts a coronal scale height of $0.194 R_{\star}$ for EQ Peg A and $0.216 R_{\star}$ for EQ Peg B where $R_{\star}$ is the radius of the respective star. 

Table \ref{tbl:eqpeg} provides a summary of stellar properties and information derived by Zeeman Doppler Imaging by \citet{Morin2008} showing the differences between the two stars.
The magnetic energy components are broken down using spherical harmonic decomposition for each star.
The sum of these decomposition components is used to normalize the magnetic field to the average magnetic field for each star and model the radial profile of the field; this is discussed in more detail in section 7. 
\citet{Morin2008} also found that the ZDI map of EQ Peg A had a strong magnetic field spot of 0.8 kG while EQ Peg B had a strong spot of 1.2 kG. 

\begin{deluxetable}{c c c c c c c c c c c c c}	
\rotate{}
\tabletypesize{\footnotesize}
\tablecaption{EQ Peg A and EQ Peg B Physical and Magnetic Properties}
\tablecolumns{8}
\tablewidth{0pt}
\tablehead{
\vspace{ -1.21cm}  
}
\startdata
	Star & Spectral  & Mass & Radius & Period  & Density & $\langle B \rangle$ & Poloidal  & Toroidal & Dipole & Quadrupole & Octupole & Axisymmetric \\
	& Type & (M$_{\odot}$) &  (R$_{\odot}$) &  (days)  & (10$^{10}$ cm$^{3})$ & (kG) & (\%) & (\%) & (\%) & (\%)  & (\%) & (\%) \\ \cline{1-13}
	A & M3.5 & 0.39 & 0.35 & 1.061 & 10.23$^{+7.35}_{-4.41}$ &  0.48 & 85 & 15 & 70 & 6 & 6 & 69/70\\
	B & M4.5 & 0.25 & 0.25 & 0.404 &  $<17.58$ & 0.45 & 97  & - & 79 & 8 & 5 & 92/94\\
	\enddata
\tablecomments{
The density is a corona density obtained by ZDI by \citep{Liefke2008}.
Magnetic field table values taken from \citet{Morin2008}.
$\langle B \rangle$ is the reconstructed magnetic energy and the average magnetic flux. 
Columns 7-13 list the percentage of reconstructed magnetic energy, respectively, lying in poloidal, toroidal, dipole (poloidal and $\ell$ = 1), quadrupole (poloidal and $\ell$ = 2), octupole (poloidal and $\ell$ = 3) and axisymmetric modes (m = 0, m  $< \ell$/2.  EQ Peg A had less than 2\% of is magnetic field in higher order terms. 
\label{tbl:eqpeg}
}
\end{deluxetable}

\section{Jansky Very Large Array P-band Observations}

\begin{deluxetable}{c c c c c c c}	
\tabletypesize{\footnotesize}
\tablecaption{List of Observations}
\tablecolumns{8}
\tablewidth{0pt}
\tablehead{
\vspace{ -1.21cm}  
}
\startdata
	Observatory & Date (UTC)  & Start Time (UTC) & End Time (UTC) & Cadence (s) &  Frequency Band &  RMS \\ \cline{1-7}
	JVLA & October 14 & 03:27:00 & 08:26:10 & 2 & P & 115.0 mJy \\
	JVLA & October 15 & 02:31:00 & 07:26:10 & 2 & P & 305.27 mJy \\
	JVLA & October 16 & 06:20:00 & 07:22:14 & 2 & P & - \\
	JVLA & October 17 & 02:19:08 & 07:18:18 & 2 & P & 122.22 mJy \\ \cline{1-7}
	APO & October 14 & 03:15:45 & 08:46:58 & 45 & u' & 0.04 mag\\
	APO & October 15 & 01:31:30 & 07:47:54 & 30 & u' & 0.05 mag\\ 
	APO & October 16 & 01:33:43 & 06:12:43 & 25 & u' & 0.04 mag \\  
	APO & October 17 & 01:27:32 & 07:41:47 & 25 & u' & 0.05 mag\\
	\enddata
\label{tbl:octobs}
\end{deluxetable}	

EQ Peg was observed in October 2015 (project 15B-291) using the JVLA's P-band receiver (230-470 MHz) in D configuration.  
Table \ref{tbl:octobs} summarizes the specifics for the 4 nights of observations.
The JVLA's P-band receiver is broken into 16 spectral windows, each comprised of 128 channels, with channel width of 125 kHz.  
The integration time for these observations was 2 seconds.  
The D configuration is the most compact configuration with an approximate angular resolution of 300 arcsec\footnote{https://science.nrao.edu/facilities/vla/docs/manuals/oss/performance/resolution} after a factor of 1.5 is added due to the use of a natural weight map.
This is higher than the separation of the EQ Peg binary system and thus it cannot resolve out the two stars.  
The NRAO VLA Sky Survey \citep{Condon1998} lists only one source within the beam, but its flux is below the confusion limit.  
D configuration is most susceptible to radio frequency interference (RFI) and the theoretical sensitivity is confusion limited for Stokes I, but still thermal noise limited for Stokes V. 
For the full observation at 370 MHz, the theoretical confusion limit was $\approx$9.7 mJy/beam\footnote{https://obs.vla.nrao.edu/ect/} using the online exposure time calculations.  
This calculation does not include the Galactic sky background which begins to dominate the system temperature at low Galactic latitudes\footnote{https://science.nrao.edu/facilities/vla/docs/manuals/propvla/determining}.
EQ Peg's Galactic Latitude is (98.5758$^{\circ}$, -39.1419$^{\circ}$).

Each observation interval lasted for five hours and was performed during the night to reduce the potential RFI.
Observations consisted of first observing the flux calibrator (quasar 3C48) in order to calibrate the bandpass and the delay calibrations.  
Then the telescope alternated between a phase calibrator (quasar J2254+2445) and EQ Peg, with approximately 10 minutes on each EQ Peg scan.  
The phase calibrator was used to calibrate phase and gain changes over time.  
Polarization calibration was also performed, but at the time of these observations the JVLA's P-band calibrations were still poorly understood.  
Therefore, the derived values for Stokes Q, U, and V should be treated with caution.   
Only the Stokes I and V components are compared in this paper.  

The data is averaged into bins of 15s in time and 500 kHz in frequency.
The theoretical sensitivity for this interval at 370 MHz is 33 mJy, after a 1.2 factor decrease in time is added to help account for the medium galactic latitude of EQ Peg.
The Galactic Sky background contributes more to the system temperature at lower latitudes, it is expected that the observational sensitivities will be larger than the ideal case.

\begin{figure}
\centering
\includegraphics[angle=-90,scale=.5,clip=true, trim=0cm 0cm 0cm 0cm]{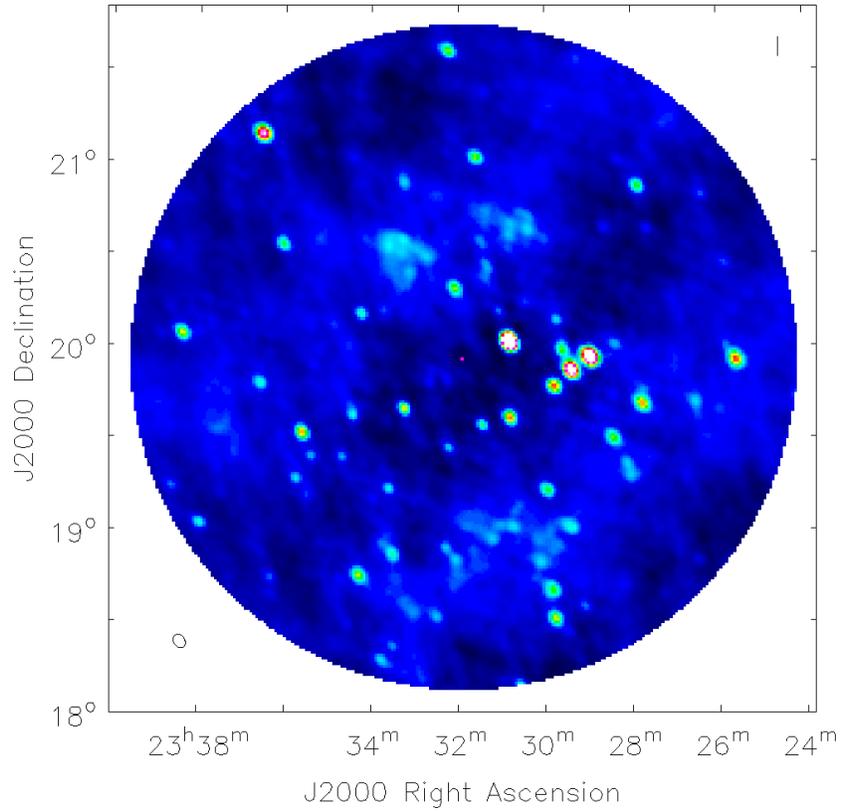} 
\caption{
Stokes I sky images of the October 14th observations.
The pink dot in the center is the coordinates of EQ Peg.
The restoring beam is shown in the lower left. 
The intensity has been scaled to show more of the background sources. 
}
\label{fig:sky}
\end{figure}

The initial clean image is used for phase self-calibration. 
Figure \ref{fig:sky} is an example of the self calibrated sky image of the October 14th observations. 
The pink dot is the coordinates of the expected location of EQ Peg (23$^{\rm h}$ $31^{\rm m}$ $52.89^{\rm s}$, +19$^{\circ}$ 56' 13.06''), after correcting for proper motion \citep{vanLeeuwen2007}, as no emission was detected from EQ Peg during this observation.
The self-calibrated full-observation image has all background sources modeled to the confusion limit threshold.  
The modeled background source visibilities are subtracted from the visibility data set leaving only visibilities from the desired region of sky.  

A dynamic spectrum for both Stokes I and V is created by calculating the flux at each point in time and frequency.  
The P band receiver contains the 4 polarization correlations XX, XY, YX, and YY from the linear feeds.  
Stokes I is defined as $(XX+YY)/2$ and Stokes V is $(XY-YX)/2i$ \footnote{https://science.nrao.edu/science/meetings/2014/14th-synthesis-imaging-workshop/lectures-files/MoellenbrockCalibration2014\_FINAL.pdf}.
A final round of flagging was performed after creation of the dynamic spectra and identification of blocks of time and frequency that were anomalously high.  
For each frequency channel, a 3rd degree polynomial is fitted to the intensity over time and any channels with a root-mean-square (RMS) of 1.5 times the median value are flagged.  

As Table \ref{tbl:octobs} indicates, three of the nights produced data sets which could be fully calibrated and dynamic spectra created.
October 16th only collected 62 of the 300 minutes scheduled for observations.  
The absence of most of the scheduled observation makes calibration problematic and therefore unable to generate a sufficiently resolved image.
Additionally, there is no overlap with the optical observations for this particular night, so we discard it. 

\section{Apache Point Observatory Optical Observations}

The APO was used to take ground-based optical observations, for the purpose of monitoring flare activity, of EQ Peg during the same time period in which radio observations were taking place at the JVLA.  
The 0.5-meter Astrophysical Research Consortium Small Aperture Telescope (ARCSAT)  performed the measurements using its University of Washington Flare Camera (FlareCam) and a Sloan Digital Sky Survey (SDSS) u' filter.
The FlareCam has a field size of $1024\times1024$ pixels which corresponds to a FOV of 11.2 arcmin $\times$ 11.2 arcmin producing a pixel scale of 0.656 arcsec/pix.\footnote{https://www.apo.nmsu.edu/Telescopes/ARCSAT/Instruments/arcsat\_instruments.html}
The SDSS u' filter has a bandpass centered at 3557\AA$ $ with a 599\AA$ $ bandwidth \citep{Fukugita1996}. 
 
The observation procedure consisted of calibration imaging, focusing the telescope lens on a nearby bright star, then performing repeated observations on the target star EQ Peg.   
The EQ Peg binary was not resolved and thus the system brightness was measured. 
The observing cadence varied between nights, between 25 and 45s, such that each time bin of the light curve would have a signal-to-noise ratio (SNR) of 3 or better after accounting for changes in atmospheric conditions.  
Table \ref{tbl:octobs} lists the specifics for each observation.
Calibration images were repeated several times and the median values per pixel were chosen.  
Firstly, sky or dome flats were observed with the intention to compensate for imperfections in the light path.  The camera is exposed to a uniform light so that pixel response can be mapped.  
Next, instant exposure bias frames are taken to compensate for read-out noise and interference from the computer.  
Finally, dark frames, at the same duration as the science target observations, are observed to compensate for the thermal properties of the charge coupled device (CCD) chip.  
The final observation is described in a dimensionless manner by:
\begin{equation}
Final = \frac{(Raw - Bias - Dark)}{Flat} \; .
\end{equation}	
This provides images that best represent the actual image of the field without the imperfections and noise of the telescope camera.  

The final observations are then transformed into a dimensionless differential light curve via:
\begin{equation}
Data = \frac{(Target - Background)}{\sum_{i} (Reference_{i} - Background_{i})}\; .
\end{equation}
The background for a given star is defined as the average background in a large blank portion of sky which is then scaled to the size of that star.  
The target is EQ Peg and Reference are all other persistent nearby stars.  

EQ Peg has an observed $U$ band magnitude of 12.737 \citep{Koen2010}.
\citet{Hilton2011} looked at how Johnson U band magnitude compared to the SDSS u-band magnitude for flares on EV Lac and showed that the flare energies are approximately equivalent; we will also treat these two as equivalent for EQ Peg.  
The median of the differential light curve is set to match this value and then subtracted to make a $-\Delta$magnitude curve.
This shows how the light changes from its quiescent value.
The negative makes increasing brightness a more positive value making it more visually intuitive.

Flares present in these curves have their energy determined by first measuring the quiescent luminosity of the star via:
\begin{equation}
L_{u,q} =  f_{q}(4\pi d^{2})\Delta \nu 
\end{equation}	
where $f_{q}$ is the u-band quiescent flux density level in units of erg/cm$^{2}$/s/Hz, $d$ is the distance to the star, and $\Delta \nu$ is the frequency bandwidth.  
The observfation corresponding to the identified flares are integrated in time in a dimensionless manner and converted into energy via:
\begin{equation}
E_{u} = L_{u,q}  \int_{t}^{t'} \frac{f(t) - f_{q}}{f_{q}} dt
\end{equation}	
where $f(t)$ is the value of the raw data after the median value of the observation has been set to the reference value of $f_{q}$.
Flare duration is defined as the time difference between initial brightening and the point where the flare has decayed to half of the maximum flux.  
The RMS of the quiescent measurements is determined by the Gaussian spread in deviations from a high ranking polynomial fit to the data points not associated to a flare event.  
The RMS values are listed in Table \ref{tbl:octobs}.  

\section{Observational Results}

\begin{figure}
\centering
\includegraphics[angle=0,scale=1,clip=true, trim=1.65cm 8cm 0cm 10cm]{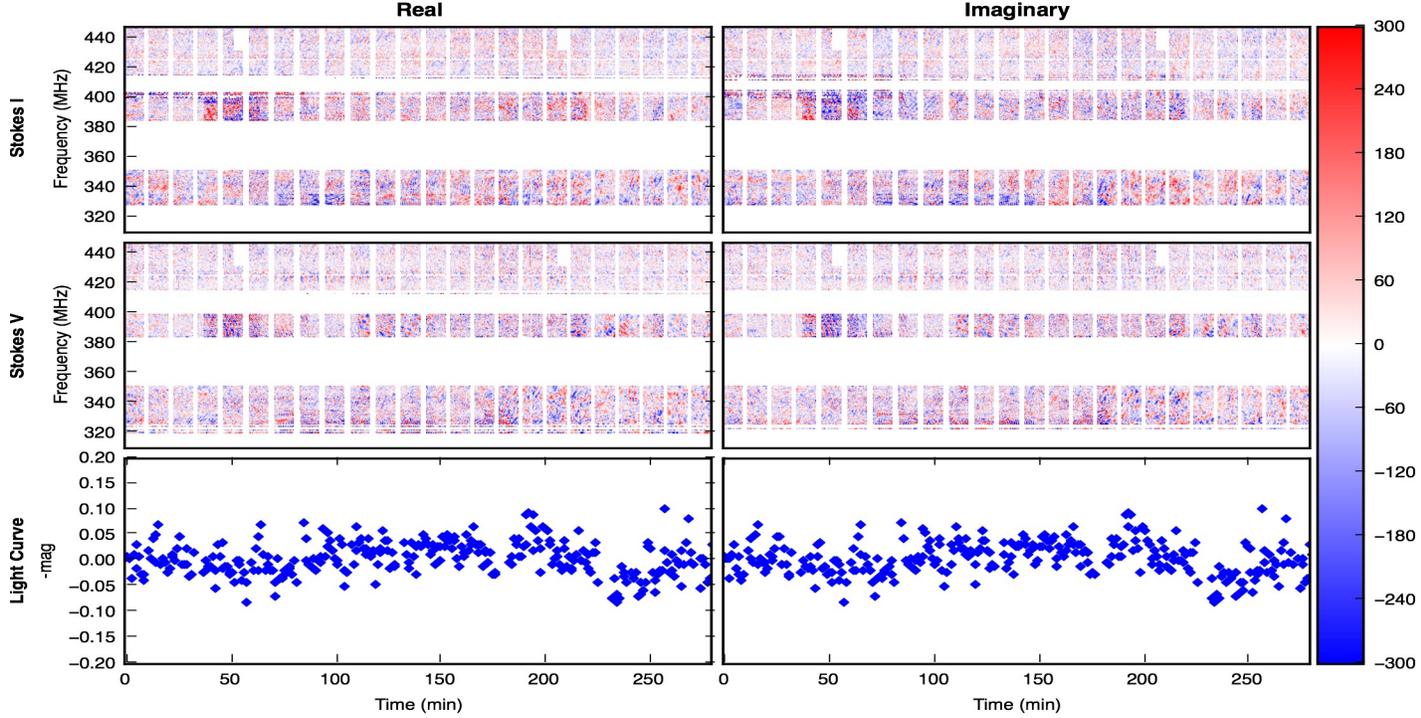} 
\caption{
Observations on October 14th.
No magnetic activity is apparent, either in flares at optical wavelengths or any variable low frequency radio emission.
The top left image is the real component of Stokes I, while the top right is the imaginary component. 
The middle two images are the real and imaginary components of Stokes V, respectively. 
The bottom plot, repeated on the right side, is the APO optical flare light curve.  
The time axis spans the entire radio observation.  
The gaps are either from a calibrator scan (vertical gaps) or due to RFI removal (horizontal blocks).  
Cyan dots on the light curve represent points with a signal-to-noise less than 3. 
The color bar indicates the range of flux density values in the dynamic spectrum in units of mJy/beam.
The median RMS for the real component of Stokes I dynamic spectra is 115.0 mJy.
Each pixel of the dynamic spectra covers $\Delta \nu =$ 500 kHz and a $\Delta t_{ds} = $15 s.
The time binning of the light curve is $\Delta t_{lc} = 45 s$.  
No flares were observed.
}
\label{fig:o14}
\end{figure}

\begin{figure}
\centering
\includegraphics[angle=0,scale=1,clip=true, trim=1.65cm 8cm 0cm 10cm]{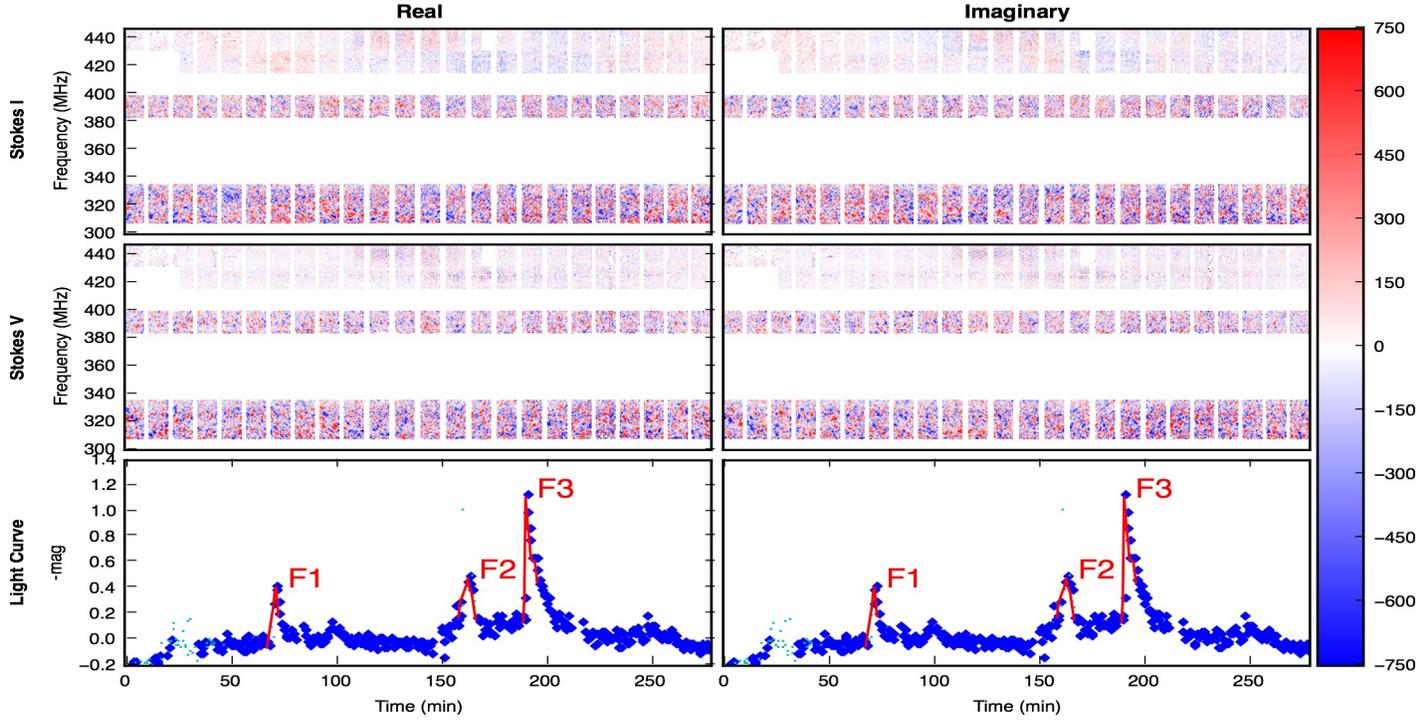} 
\caption{
Observations on October 15th.
The median RMS for the real component of Stokes I dynamic spectra is 305.27 mJy/beam. 
Red curves display flare points used to calculate flare energies.
Three flares are observed, and labeled F1, F2, and F3.
From left to right, their flare energies are $E_{u}$ = $0.9\times10^{31}$, $1.5\times10^{31}$, and $5.8\times10^{31}$ erg. 
The time binning of the light curve is $\Delta t_{lc} = 30 s$.  
Other details are the same as in Figure \ref{fig:o14}.
}
\label{fig:o15}
\end{figure}

\begin{figure}
\centering
\includegraphics[angle=0,scale=1,clip=true, trim=1.65cm 8cm 0cm 10cm]{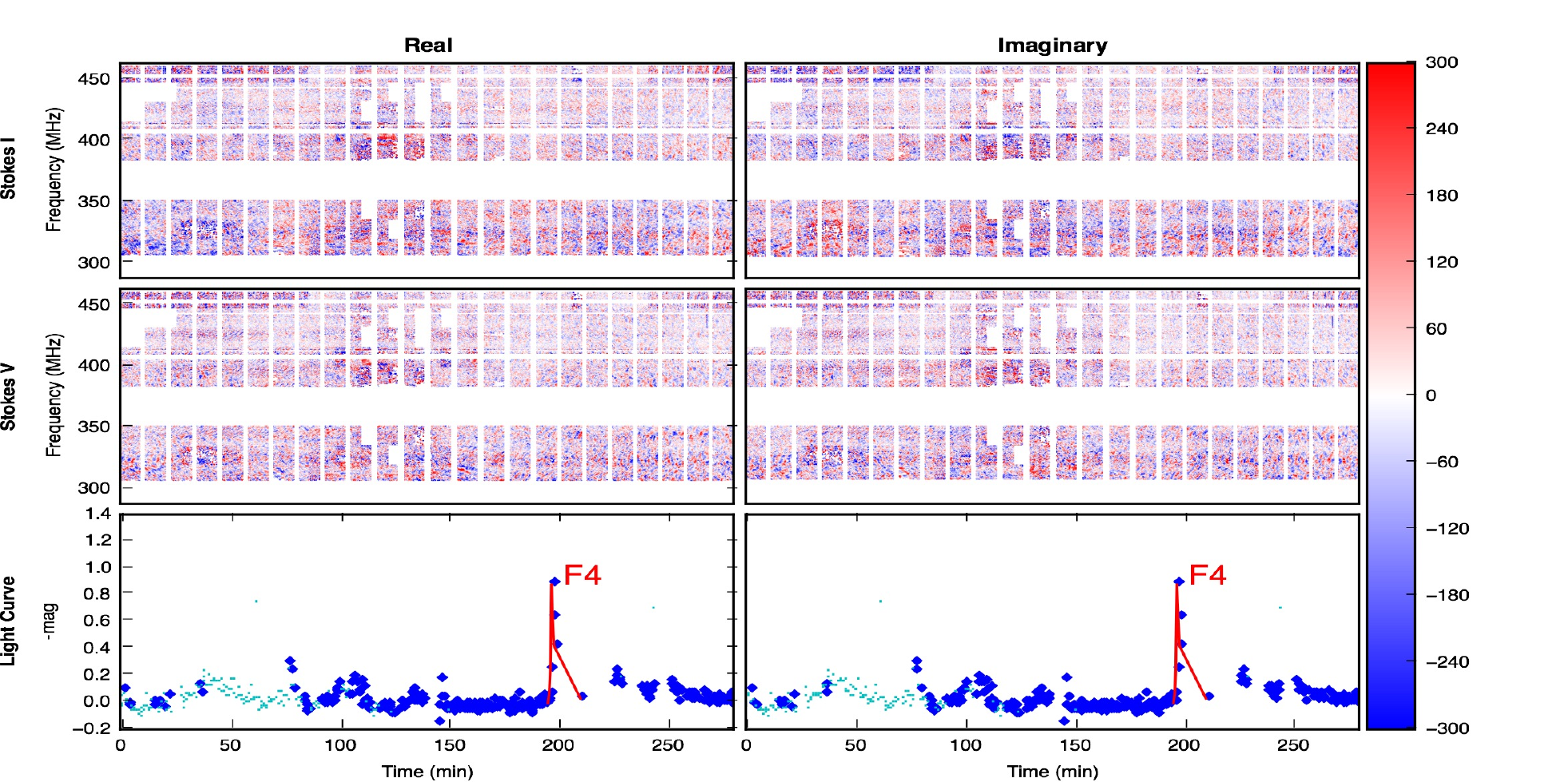} 
\caption{
Observations on October 17th. 
The median RMS for the real component of Stokes I dynamic spectra is 122.22 mJy/beam.
Red curves display flare points used to calculate the flare energy.
The flare, labeled F4, had an energy of $1.4\times10^{31}$ erg.  
The time binning of the light curve is $\Delta t_{lc} = 25 s$.  
Other details are the same as in Figure \ref{fig:o14}.
}
\label{fig:o17}
\end{figure}

Figures \ref{fig:o14}, \ref{fig:o15}, and \ref{fig:o17} show the dynamic spectra for three nights.
Each figure shows the real and imaginary component of the Stokes I and V components.  
The bottom panels also show the APO flare light curve for the observation.  
Flares points used to calculate the flare energies are traced in red. 
Cyan points in the flare curve represent points below a SNR of 3. 
The night of October 16th contained no flares and did not properly overlap with the radio observations.  
Also, the radio observations had RFI issues rendering accurate calibration problematic.
For these reasons, this night has been omitted.  
The color bar indicates the range of flux density values in the dynamic spectrum in units of mJy/beam.
Each pixel of the dynamic spectra covers $\Delta \nu =$ 500kHz and a $\Delta t_{ds} = $15 s which corresponds to a theoretical sensitivity of $\sim$30 mJy/beam.
The observational median Stokes I RMS sensitivities range from 100-300 mJy/beam are listed in Table \ref{tbl:octobs}.
Events such as ionospheric disturbances could increase our observational sensitivity above the theoretical limit as well as potential confusion from sources outside the primary beam. 
Therefore any events would need to be a few hundred mJy to be detected at a significance of a few sigma. 

\section{Methodology for Interpreting Stellar CME-Flare Data}

The following sections provide an abbreviated version of the multi-wavelength analysis to describe the use of type II radio bursts and optical flare measurements to characterize CME's. See \citet{Crosley2016} for full details. 

\subsection{Radio Burst Observations}

A type II burst is an easily identifiably signal unique to CME's.
A MHD shock will be produced when the CME is traveling with sufficient velocity through the stellar atmosphere.
This frequency changes as the source travels through the atmosphere producing a distinct slope in time and frequency.
An outward moving shock will emit at a lower frequencies as it travels towards lower densities.
The frequency will vary in time as:
\begin{equation}
\frac{d\nu}{dt}=\frac{\partial\nu}{\partial n}\frac{\partial n}{\partial h}\frac{\partial h}{\partial s}\frac{\partial s}{\partial t} 
\end{equation}
where $\nu$ is frequency, $n$ is electron density, $h$ is radial height above the star, $s$ is distance along the path which the shock travels, and $t$ is time.  

The drift rate $\left({d\nu}/{dt}\right)$ is composed of four terms describing the changing environment around the shock.  
Type IIs radiate radio emission at the fundamental and harmonic of the local plasma frequency ($\nu_{p} \sim n^{1/2})$ \citep{Gopalswamy2006}; thus the first term describes how the emitted frequency changes with local density $\left({\partial \nu}/{\partial n} = {\nu}/{2n} \right)$.  
A barometric model $\left({\partial n}/{\partial h} = {- n}/{H} \right)$ of the stellar atmosphere is used where $H$ describes how the density changes with height $(n(h) = n_{0}e^{-h/H})$. 
Equation (\ref{equation:scaleheight}) is used to determine the scale height for each star and the density at the base of the corona ($h=0$) is normalized to the densities listed in Table \ref{tbl:eqpeg}.   
The path describing how the CME moves through the stellar atmosphere is assumed to be a straight line $\left({\partial h}/{\partial s} = \cos \theta \right)$ which travels orthogonally to the stellar surface ($\cos \theta = 1)$.
The distances examined are sufficiently small such that any acceleration of the CME is negligible and therefore the fourth term describing the velocity of the shock is a constant $\left({ds}/{dt} = v_{s}\right)$.   

Substituting all the differential terms and rearranging leads to the shock speed expression:
\begin{equation}
\label{eq:shockspeed}
v_{s} = \frac{-2H\dot{\nu}}{\nu}
\end{equation}
where $\nu$ is frequency, $\dot{\nu}$ is the frequency drift rate,  $v_s$ is velocity of the shock, and $H$ is the density scale height.
The emitted frequency $(\nu)$ and drift rate $(\dot{\nu})$ of the type II burst are directly measured in observations while the coronal scale height $H$ is modeled.
This leaves the shock velocity completely constrained by the other parameters.  

A detection of a drifting radio burst provides a constraint on the speed of the CME shock and therefore limits the speed of the CME itself.   
The uniqueness of type II bursts to CME's allows a measurement of type II burst occurrence rate to be a lower bound for the occurrence rate of CME's as not every CME produces a type II burst.

A sweeping timescale ($\tau$) can be defined from equation (\ref{eq:shockspeed}) as
\begin{equation}
\frac{\dot{\nu}}{\nu} = \frac{-v_{s}}{2H} \rightarrow \nu = e^{\frac{-v_{s}}{2H}t} = e^{-t/\tau} \rightarrow \tau = \frac{2H}{v_{s}}\; .
\end{equation}
Using an assumed shock speed of 3000km/s this translates to a sweeping time required for the frequency to change by a factor of 1/e of 31 s for EQ Peg A and 25 s for EQ Peg B.  
The frequencies for the JVLA P-band range over roughly a factor of 2 leading to only 75\% of the sweeping time being required to pass through the entire band range.
This means an expected burst for EQ Peg A would last 23s and it would last 19s on EQ Peg B. 

\subsection{Flare Observations}

\citet{Yashiro2006} has shown, using GOES X-ray measurements, that solar CME's have an increasing association rate to solar flares as flare flux, flare peak flux, and flare duration increase.
The association rate saturates to 100\% when the flare is sufficiently large.  
Building from this connection, \citet{Emslie2005}, \citet{Emslie2012} and \citet{Osten2015} have shown evidence supporting an equipartition between the total radiated energy of the flare and the mechanical energy of its associated CME. 
This relation can be described as:  
\begin{equation}
\label{eq:equipar}
\frac{1}{2} M_{CME} v^{2} = \frac{E_{rad}}{\epsilon f_{rad}} 
\end{equation}
where $f_{rad}$ is the fraction of the bolometric radiated flare energy appropriate for the waveband in which the energy of the flare is being measured \citep{Osten2015}.
The factor $\epsilon$ is a constant of proportionality $\approx0.3$ to describe the relationship between bolometric radiated energy and CME kinetic energy.  
The velocity of the CME is $v$, its total mass is $M_{CME}$, and the radiated flare energy is $E_{rad}$.

\citet{Aarnio2012} and \citet{Drake2013} have found an empirical relationship between the solar flare X-ray energy and its associated CME mass.
It is a relationship of the form:
\begin{equation}
\label{eq:Mcme}
M_{CME} = A E^{ \gamma} \mbox{  [g]}
\end{equation}
where $M_{CME}$ is the CME mass, A is a constant of proportionality, and $\gamma$ is the power law index.  
\citet{Drake2013} lists $A = 10^{-1.5 \mp 0.5} $ in cgs units and $\gamma = 0.59 \pm 0.02$.   

Assuming the definition of $M_{CME}$ is mutually consistent between formulas equation. (\ref{eq:equipar}) and (\ref{eq:Mcme}), they can be utilized in concert to solve for either the flare energy as a function of CME velocity:
\begin{equation}
\label{eq:ECombined}
E_{rad} = \left[ \frac{A\epsilon v^{2}}{2} f_{rad}  \right]^{\frac{1}{1-\gamma}}
\end{equation}
or the velocity as a function of flare energy:
\begin{equation}
\label{eq:VCombined}
v = \sqrt{\frac{2}{A\epsilon f_{rad}}}(E_{rad})^{\frac{1-\gamma}{2}}\; .
\end{equation}
These are now a single variable equations of $E_{rad}(v)$ and $v(E_{rad})$.  
The errors of the above equations are discussed in detail in \citet{Crosley2017}.  
The results of equation (\ref{eq:ECombined}) is suggested to be only useful to an approximate order of magnitude while equation (\ref{eq:VCombined}) is suggested to be accurate to within a factor of 2.

\section{Discussion of Results Utilizing Interpretation Methodology}
\subsection{Flare Analysis and Event Modeling}

The nights on the 15th and 17th of October contained flares, Table 3 summarizes their properties. 
On the 15th,  there were three flares with emitted energies of 0.9, 1.5, and 5.8 $\times10^{31}$ erg.  
If CME's accompanied these flares as anticipated, equation (\ref{eq:VCombined}) evaluates the flare energies and infers that the associated CMEs have speeds of 1315, 1451, and 1925 km/s respectively. 
On the 17th, there was only 1 flare with an emitted energy of $1.4 \times10^{31}$; corresponding to a velocity of 1431 km/s.
Table 3 lists a summary of these results.  
Taking the u-band energy at which CME's have a 100\% correlation to flares from \citet{Yashiro2006} and converting to the GOES band $\left(f_{u}/f_{GOES} = {0.06}/{0.11} \mbox{ \citep{Osten2015}}\right)$ is $ E_{u}$ = 8.38$\times10^{29}$ erg.
Therefore, all of these events should have produced a CME if the saturation limit for a 100\% association of flares to CMEs occurs at the same flare energy as the Sun.

\begin{deluxetable}{c c c c c}	
\tabletypesize{\footnotesize}
\tablecaption{Observed Flare Properties}
\tablecolumns{8}
\tablewidth{0pt}
\tablehead{
\vspace{ -1.21cm}  
}
\startdata
	Flare Label & Date  & $\Delta mag_{peak}$ & E$_{total}$  & Estimated V$_{CME}$  \\
	& {(UTC)} &  & (10$^{31}$ erg)  & (km/s)\\ \cline{1-5}
	F1 & Oct. 15th & 0.41 $\pm$ 0.05 & 0.9 $\pm$ 0.18 & 1315 $\pm$ 659 \\
	F2 & Oct. 15th & 0.48 $\pm$ 0.05 & 1.5 $\pm$ 0.23 & 1451 $\pm$ 727 \\
	F3 & Oct. 15th & 1.11 $\pm$ 0.05 & 5.8 $\pm$ 0.46 & 1925 $\pm$ 964 \\
	F4 & Oct. 17th & 0.88 $\pm$ 0.05 & 1.4 $\pm$ 0.18  & 1431 $\pm$ 717 \\
	\enddata
\tablecomments{Summary of the observed flares.
Inferred CME velocity determined by equation (\ref{eq:VCombined}).
\label{tbl:flare}
}
\end{deluxetable}


For these associated CME speeds to be able to drive a MHD shock required to produce the type II radio burst, they need to be faster than the Alfv\'{e}n speed $\left(v_{A} = {B}/{\sqrt{\mu_{0}m_{e}n}}\right)$.
The calculation of the Alfv\'{e}n speed is only valid up until a few stellar radii out from the star where stellar wind begins to dominate the magnetic field.  
On the sun, this breaks down at about 5 $R_{\odot}$ \citep{Bravo1998} while \citet{Vidotto2014} used 4 $R_{\star}$ when modeling several M dwarfs. 

We determine the Alfv\'{e}n speeds as a function of R$_{\star}$  for each star using each stars' corresponding modeled magnetic field and barometric model.  
The magnetic field properties for EQ Peg A and B are listed in Table \ref{tbl:eqpeg}.  
The components of the magnetic energy are broken down with spherical harmonic decomposition ($\ell$ = 0, 1, 2, etc.).  
The portion of magnetic energy contained in the first three terms, that is the dipole (poloidal and $\ell$ =1), quadrupole, (poloidal and $\ell$ = 2), and octuple (poloidal and $\ell$ = 3) are used to reconstruct the magnetic field.  
Table \ref{tbl:eqpeg} states that both EQ Peg A and EQ Peg B are highly axisymmetric and thus we will only model the simplest m=0 terms for the spherical harmonics.  
Therefore the magnetic field equation will take the form:
\begin{equation}
B(r,\theta) = N\left(\frac{a}{r^{3}}Y_{1}^{0}(\theta) + \frac{b}{r^{4}}Y_{2}^{0}(\theta) + \frac{c}{r^{5}}Y_{3}^{0}(\theta)\right)
\end{equation}
where $N$ is the normalization factor, $Y_{\ell}^{m}$ is the spherical harmonic, and the constants are the percentage of magnetic energy contained in their respective modes listed in Table \ref{tbl:eqpeg}.
We then average $B(r,\theta)$ over $\theta$ from 0 to $\pi/2$ to determine a simplified $B(r)$.

For EQ Peg A, this takes the final form 
\begin{equation}
B(r) = \frac{480}{2.93} \left[  \frac{0.7}{r^{3}} \sqrt{\frac{3}{\pi^{3}}}+ \frac{0.06}{r^{4}} \left(\frac{1}{8}\sqrt{\frac{5}{\pi}}\right)+ \frac{0.06}{r^{5}} \left(\frac{1}{6}\sqrt{\frac{7}{\pi^{3}}}\right)\right]
\end{equation}
and for EQ Peg B it is 
\begin{equation}
B(r) = \frac{450}{3.31} \left[  \frac{0.79}{r^{3}} \sqrt{\frac{3}{\pi^{3}}}+ \frac{0.08}{r^{4}} \left(\frac{1}{8}\sqrt{\frac{5}{\pi}}\right)+ \frac{0.05}{r^{5}} \left(\frac{1}{6}\sqrt{\frac{7}{\pi^{3}}}\right)\right]
\end{equation}
where $r$ is in units of stellar radius and the magnetic field is in units of G.
The normalization factor is made such that the magnetic field has a value equal to the average magnetic field listed in Table \ref{tbl:eqpeg} at the base of the corona ($r = R_{\star}$).

Figure \ref{fig:VA} displays the CME velocities determined from these flare energies, shown in black, as compared to the Alfv\'{e}n speed as a function of distance for EQ Peg A and B.  
The magenta shaded region represent the error from these values listed in Table \ref{tbl:flare}.
The red line represents the Alfv\'{e}n speed $\left(v_{A} = {B}/{\sqrt{\mu_{0}m_{e}n}}\right)$,  limit for an event originating from EQ Peg A which would be observable by the JVLA's P-band (230-470 MHz).
The position of the red line is determined by the frequency which is emitted ($\nu \sim \sqrt{n}$) at the distance as determined by the barometric model.
For a shock to have occurred, the flare determined velocities (black lines), would need to be above the red line. 
For EQ Peg A, the upper red dotted line, corresponds to a decreased base density which pushes the associated distances closer.  
The blue line is the lower limit for the Alfv\'{e}n speed for EQ Peg B as its density is an upper limit. 
EQ Peg B has a higher measured density which increases the distance from the base of corona required to emit in the desired frequencies.  
The distance above the stellar surface is scaled to each stars' respective R$_{\star}$.  

The flare-estimated CME velocities all tend to be below the expected Alfv\'{e}n speed limit at all distances for EQ Peg A and there is some distance at which an event would have potentially burst on EQ Peg B.   
However, this does not rule out all shocks from having occurred. 
The event would likely be required to be emitted in a high density region for the shock to have formed.  
From Figure \ref{fig:VA}, an event observable in the P-band data range spans distances of about 0.3 R$_{\star}$ for each star.  
Traversing this distance using the averaged predicted velocity from the four flares (1530 km/s) would take 48 and 34 seconds to completely traverse this window for EQ Peg A and B respectively.  
Any signals present would have a theoretical drift rate through the JVLA's P-band of 5.03 MHz/s and 7.04 MHz/s if they occurred on EQ Peg A and EQ Peg B respectively.  
Since EQ Peg B has a smaller radius,  this leads to a shorter time to traverse 0.3 R$_{\star}$ and through the observable frequency range leading to a faster drift rate than expected for EQ Peg A.  

On the scale displayed in figures \ref{fig:o14} - \ref{fig:o17}, to the eye, a burst would look very close to a vertical line in the real Stokes I, presumably ranging over the full frequency range because of the velocity, bandwidth, and time considerations discussed above.  
It should likely have a corresponding signal in the same location in the real Stokes V.  
A signal also present in the imaginary Stokes Components would indicate that it is not a real signal but an artifact of calibration or some other source.
Currently none of the dynamic spectra have this distinct feature.

\subsection{Event Frequency}

\citet{Lacy1976} performed statistical of EQ Peg flares and modeled a power law relationship describing the cumulative distribution of flare energy for EQ Peg as $\log f = \alpha + \beta \log E_{U}$. 
The units of $f$ are in $hr^{-1}$, the units of $E_{U}$ are in erg, $\alpha = 30.7 \pm 4$, and $\beta = -1 \pm 0.14$.   
The threshold energy at which a solar flare should always have an accompanying CME is 8.38 $\times10^{29}$ erg in SDSS u with an inferred flare rate of 1.2 flares per hour for the EQ Peg system.
This flare rate lowers to 0.05 flares per hour for flares larger than or equal to $0.9\times10^{31}$ erg in U, the smallest flare processed in our observations.
Over the course of the cumulative $\sim$15 hours of overlapping optical observations, we would statistically anticipate to have seen 18 flares at $8.38 \times10^{29}$ or higher, 1 of which would be above $0.9\times10^{31}$ erg \citep{Lacy1976}.
Weather was an obstacle during these observation which may have obscured events due to cloudy skies and other uncontrollable atmospheric conditions.
Four flares were observed during the 15 hours of the overlapped radio and optical time.  
The lack of activity is particularly noticeable for the night of October 14th which had no events.  
If solar scaling relations hold, the lack of powerful flares would indicate a lack of CME's and thus provide no detections.  
Similarly, October 17th only had a single flare which limits the anticipated number of events that could be observed.  
Although flares occur stochastically, the departure of the expected number of flares from observed number may stem from insufficient sampling time.

\subsection{Detectability Limits}
The plasma emission of a type II burst is coherent emission which follows the form $S_{\nu}  = {\nu^{2}}/{c^{2}}\int T_{b} d\Omega$ where c is the speed of light, the observed frequency is $\nu$, $T_{b}$ is the brightness temperature, and $d\Omega$ is the solid angle.
The solar maximum brightness temperature for a type II burst is $10^{14}$ K \citep{Benz1988}, and since M dwarfs are more magnetically active, we will assume that our stellar type II bursts $T_{b}$ will be larger.
There is evidence that radio events on stars can have higher brightness temperatures to $10^{18}$ K \citep{Osten2008}.
Therefore we will use the constraint $10^{14} < T_{b} < 10^{18} $ K.

\begin{figure}
\centering
\includegraphics[angle=-90,scale=.5,clip=true, trim=2cm 1cm 1cm 3cm]{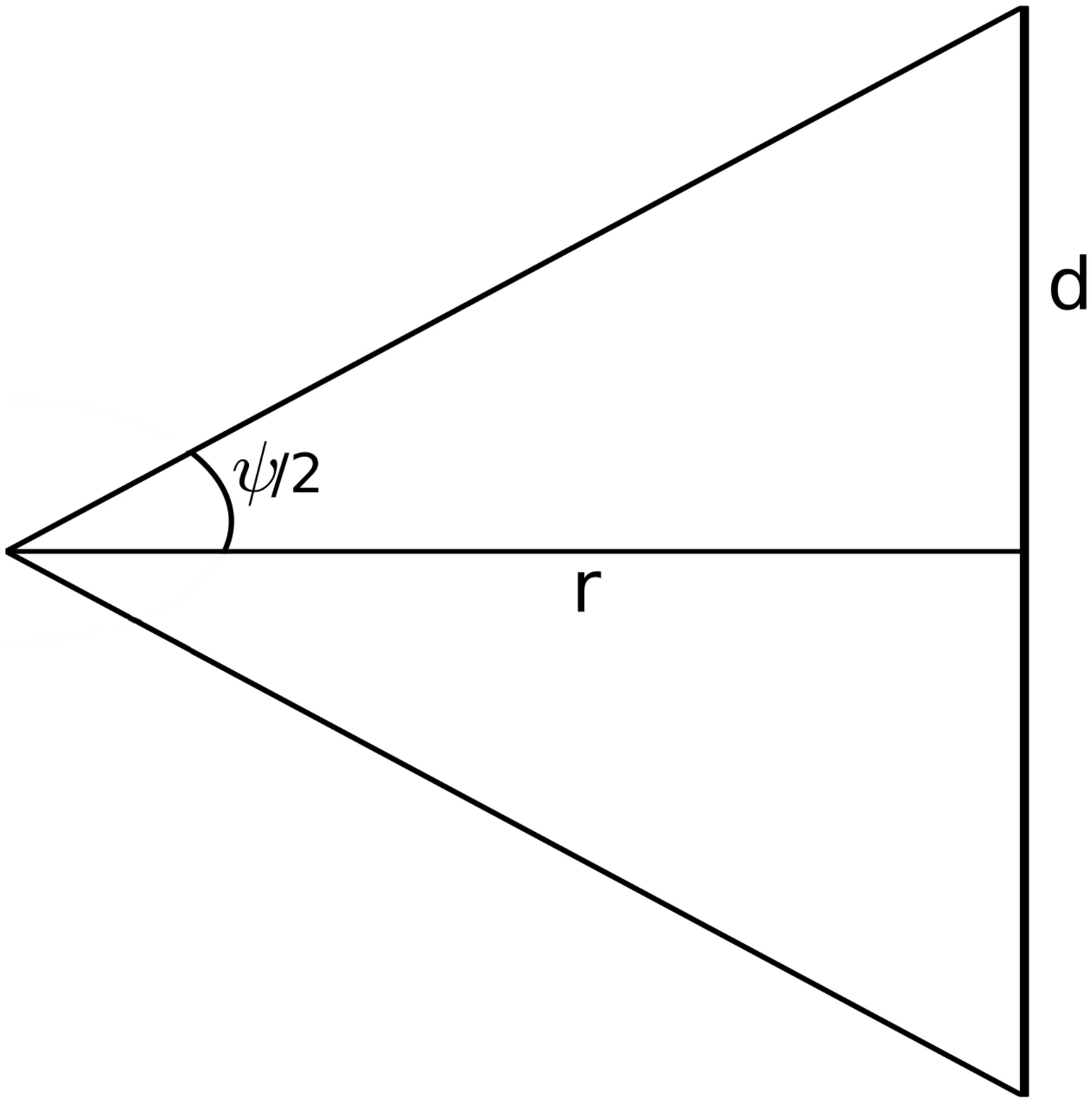} 
\caption{
Cone representing the spread of a CME as it propagates to a distance $r$ from the center of the star. 
}
\label{fig:cone}
\end{figure}

We will model the event area $\Omega$ from solar CME trends. 
\citet{Byrne2012} describes the angular spread of a CME as a function of the radial height (r) from the core of the sun as a power law expansion: $\Delta \psi(r) = \Delta \psi_{0}\left({r}/{r_{\odot}}\right)^{0.22}$, where $\Delta \psi_{0} = 26^{\circ}$.
We assume that this relationship holds for EQ Peg, $\Delta \psi(r) = 26^{\circ}\left({r}/{r_{\star}}\right)^{0.22}$.
By assuming the CME takes the shape of a cone shown in Figure \ref{fig:cone}, the base radius can be determined at a given radius r from the center of the star in units of R$_{\star}$ via: d = $r \tan{\left(\frac{26^{\circ}}{2}(\frac{r}{r_{\star}})^{0.22}\right)}$.  
Assuming the shock surface completely covers the face of the CME, the area predicted by the circle at the base of the cone ($\pi d^{2}$) predicts maximal areas between 0.53 - 1.3 R$_{\star}^{2}$ at r = 1.6 and 2.3 R$_{\star}$ respectively. 
Scaling this to the area of the star ($A_{star} = \pi R_{\star}^{2}$), states 17\% and 41\% of the stellar surface would be covered by the shock at the two maximal areas.

\begin{figure}
\centering
\includegraphics[angle=0,scale=.5,clip=true, trim=1cm 1cm 2cm 2cm]{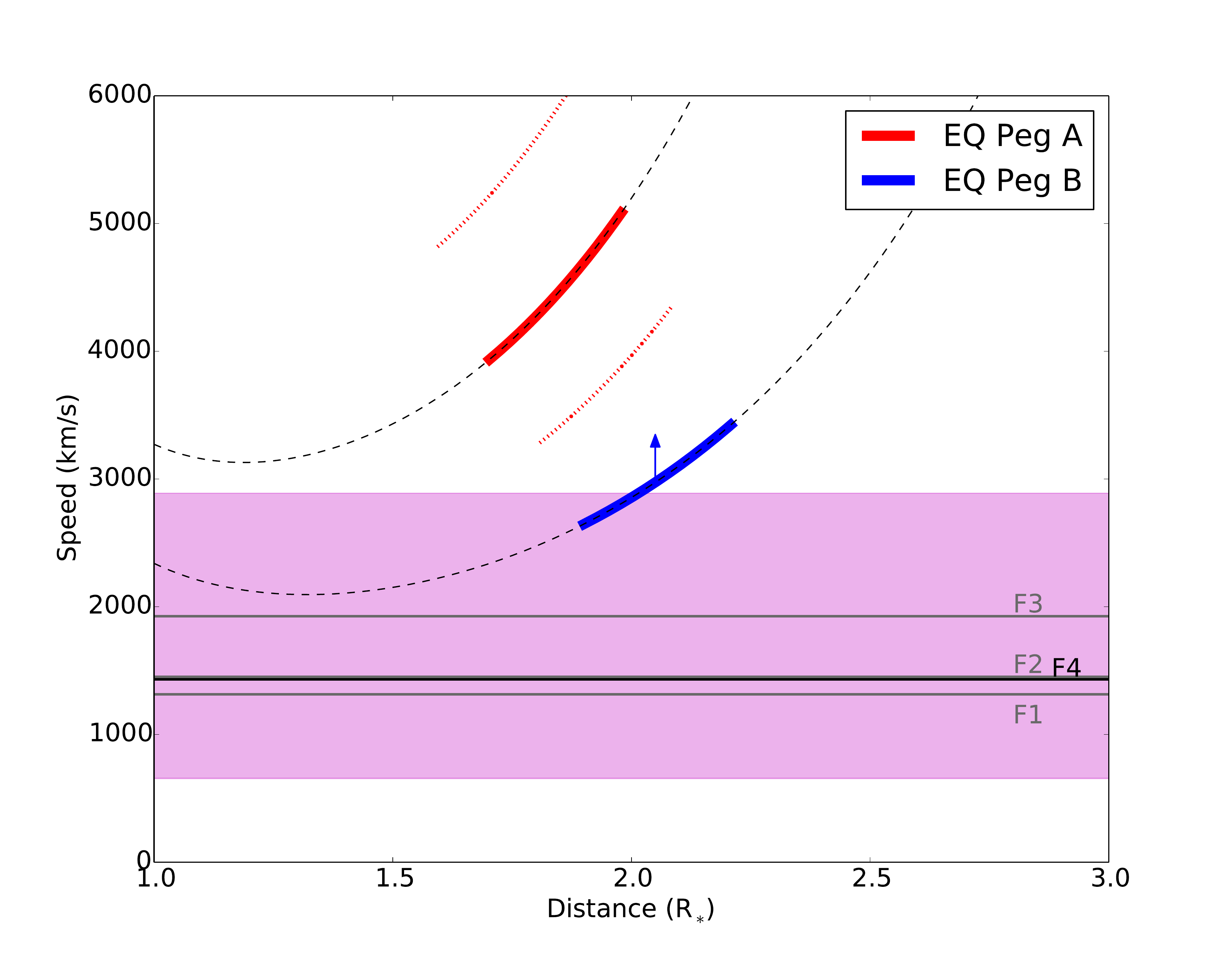} 
\caption{
Displays the flare energy determined velocities, shown in black solid lines, as compared to the Alfv\'{e}n speed as a function of distance for EQ Peg A and B.  
The dashed-black lines represent the Alfv\'{e}n speed as a function of distance.
The red line represents portion of the Alfv\'{e}n curve where an event originating from EQ Peg A which would be observable by the JVLA's P-band.
The red dashed lines are the limits after varying the density by the errors listed in Table \ref{tbl:eqpeg}.
The blue line is the same for EQ Peg B.
The blue arrow in the direction in which the Alfv\'{e}n speed resides, due to the upper limit on density, relative to the solid line. 
The magenta shaded region represent the deviation from these speeds listed in Table \ref{tbl:flare}.
}
\label{fig:VA}
\end{figure}	

Assuming a simplistic, fully random, distribution for launch angle,  the maximal shock area will be modified by $cos\theta cos\phi$ where $\theta$ is the horizontal launch angle from line of sight and $\phi$ is the vertical launch angle from line of sight.  
Using the average angle from 0 to ${\pi}/{2}$ of ${\pi}/{4}$ for both $\theta$ and $\phi$, we would expect the average shock area for an event to be 1/2 the maximum value.  
This represents the `best case scenario' for observations.  
Type II bursts are often not fully Alfv\'{e}nic and the shock will not cover the complete CME front.  
This would further lower the expected shock area for a given event. 

\begin{figure}
\centering
\includegraphics[angle=0, scale=0.8, clip=true, trim=0cm 0cm 1cm 0cm]{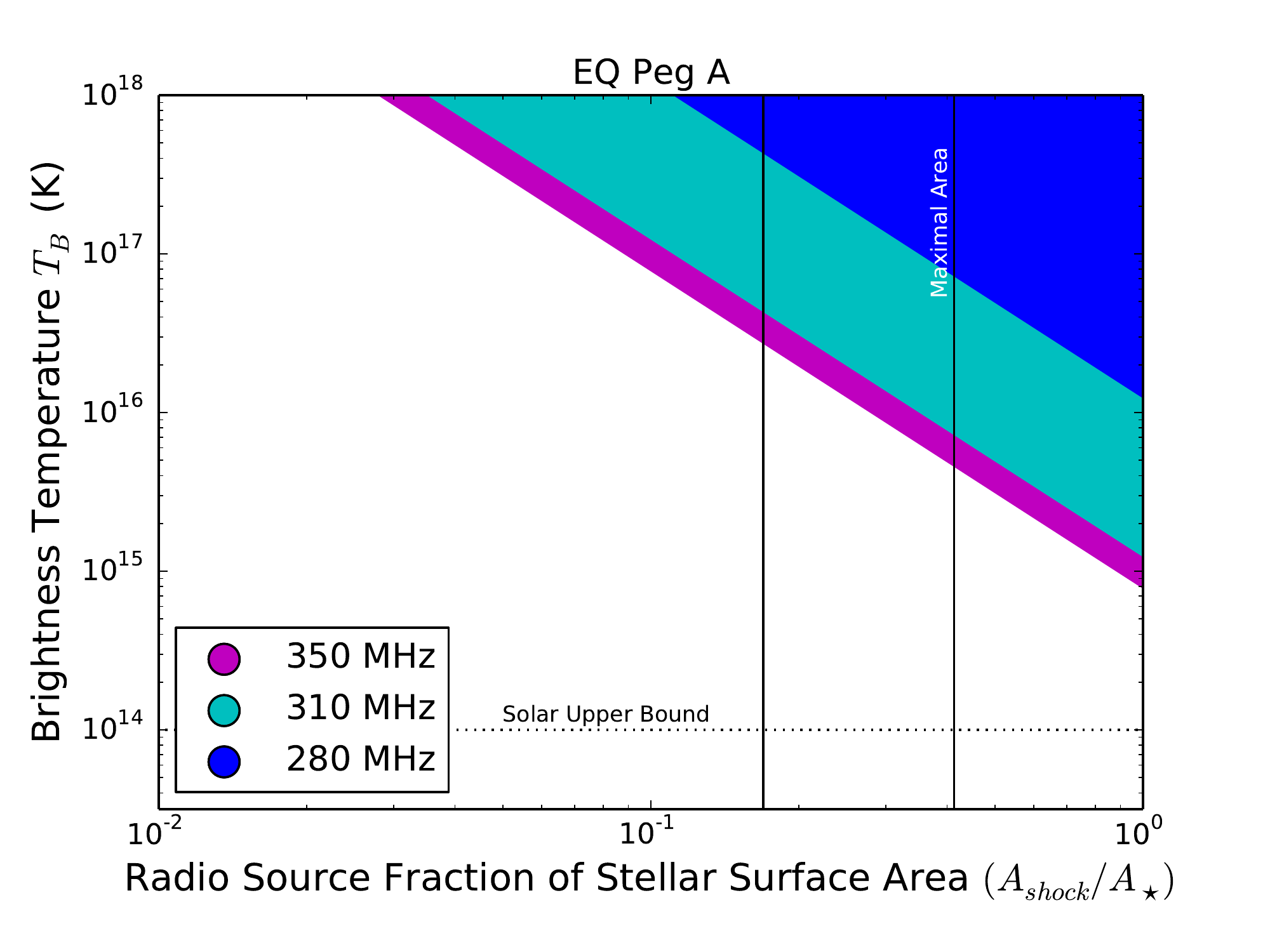} 
\caption{The theoretical detectable levels of radio emission for EQ Peg A (values are similar for EQ Peg B) at several P-band frequencies for combinations of brightness temperature and radio source extent for a 15s period over a 500 kHz bandwidth on the night of October 14.
The vertical black lines (0.17 and 0.41 $A_{\star}$) represent the maximal area expected to be observed for fully Alfv\'{e}nic shocks at 1.6 and 2.3 R$_{\star}$ respectively from left to right.
RMS for the frequencies 420, 340, and 300 are 46.06, 104.8, and 373.56 mJy.
The black dotted line represents the solar maximum brightness temperature of $10^{14}$K \citep{Osten2006}.
}
\label{fig:pband}
\end{figure}

Figure \ref{fig:pband} shows the region of parameter space constrained by the lack of detections for the night of October 14th for EQ Peg A. 
They show the detectable levels (5 times the SNR) of radio emission at several P-band frequencies for combinations of brightness temperature and radio source extent.
This figure generally applies equally for all nights to within a factor of 2.
The night of the 15th performed slightly worse while the 17th was slightly more sensitive.
An event with the combination of T$_{b}$ and source size in the shaded region would have produced flux density large enough to be detectable with the real sensitivities of our observations. 
The vertical black lines correspond to the theoretical maximal fractional surface area of 0.17 and 0.41 A$_{\star}$ that a CME would have for these events listed above. 
The black horizontal dotted line represents the solar maximum brightness temperature for a type II burst of $10^{14}$ K \citep{Benz1988}.
High frequency events which are aimed towards Earth, and thus have a large fraction of the stellar surface, require the lowest brightness temperatures to be observable.  
This makes them the best candidates for observation.   
Limb CMEs, at all frequencies, require higher brightness temperature to be detected.  
		
\section{Interpretation}

It is known that stellar flares exist and that they share many of the same properties and phenomena as solar flares suggesting a common mechanism \citep{Osten2016}.  
Studies of the solar flare - solar CME relationship \citep{Yashiro2006} have shown that they too are connected. 
Therefore, we infer that the presence and increased frequency of occurrence of solar flares should also have a relationship leading to the presence of and higher frequency of occurrence of stellar CMEs as compared to the Sun.  

The expectation is that these CMEs are able to escape and that, on occasion, the velocity of the CME is sufficient enough to drive a shock that would be analogous to the kind which produce type II bursts on the sun.  
While no type II events were detected on the stars monitored, this does not rule out them having occurred and been outside of our detection sensitivity. 

Following the logic back to the origin, what could have prevented detection?
A burst may have occurred outside of our observed frequency range, meaning the burst occurred deeper in the corona or further out, and thus will have gone undetected.  
Alternatively, a potential shock may not follow the traditional type II burst and instead emits radiation in an entirely different manner. 
Barring that however,  the coronal profile of M dwarfs may be different such that the expected frequency range of the type IIs is outside of the JVLA P-band window which was observed.
Figure \ref{fig:VA} suggests that the events would likely not have been fast enough to drive a shock at the model distances probed.
There may have been a shock that was generated at deeper in the corona and faded before reaching the observable distances.
However, the measurements on which this calculation was performed are global averages.
Smaller scale regions of the stellar atmosphere may have the right combination of density and magnetic field strength to generate a MHD shock.  
Supposing that a shock was present, it still would have needed to have a large enough combination of brightness temperature and area to be detected.  

Another possibility is that the expectation of a CME being emitted may not be true.
There is evidence that large flares on the sun do not produce CMEs in the presence of a stronger overlying arcade of fields \citep{Wang2007,Schrijver2009,Thalmann2015}.
EQ Peg, or perhaps M dwarfs in general, may have a field structure that inhibits or prevents the breakout of CMEs, so that no type II bursts would be generated.

Finally, the rate at which these events are expected to occur may be incorrect. 
\citet{Lynchprop} suggests the radio flare rates of these stars may have been overestimated due to potential RFI contamination of initial studies and the difficulty of long observation campaigns of more modern telescopes being unable to reproduce those initial flare rates \citep{Lynch2017}.  
A lower flare rate would decrease the anticipated frequency for large flares which are expected to produce CMEs.
The flare rates used here, however, are optical flare rates. 
Therefore the time scales required to observe events may be much longer than previously thought. 

There is a growing sense that the magnetic activity behavior of M dwarfs may not be a simple extrapolation from the Solar case. 
The current lack of CMEs may be indicative of more than just insufficient sensitivity.  
It is thought that frequently flaring stars may not be dominated by a constant quiescent wind but by sporadic CMEs which act as a sporadic fast and dense stellar wind.  
\citet{Wood2014} looked at young stars, including the star EV Lac which is an active M dwarf similar to the EQ Peg system, and found that EV Lac does not have particularly strong coronal winds.
The lack of CMEs goes along with this inference of weak quiescent stellar winds.  

\section{Conclusion}

Our objective was to perform a multi-wavelength analysis of EQ Peg in order to detect and constrain stellar CME's via the type II burst. 
These observations would be used to provide experimental evidence to aid in understanding of magnetic eruptive events and their impact on exoplanet habitability.  
We reported on 20 hours of observations; 5 of them were unable to be properly calibrated due to observational and calibration issues. 

The converted GOES-band energy limit for 100\% association between flares and CME's from \citet{Yashiro2006} to u-band via \citet{Osten2015} is 8.38$\times10^{29}$ erg.  
We observed four flare events with energies 0.9 $\pm$ 0.18 $\times10^{31}$, 1.5 $\pm$ 0.23$\times10^{31}$, 5.8 $\pm$ 0.18 $\times10^{31}$, and $1.4 \pm 0.18 \times10^{31}$ erg.  
If the solar scaling relationship hold, each flare should have produced a CME.
Equation (\ref{eq:VCombined}) predicts that the speeds of the associated CME's would have been 1315 $\pm$ 659, 1451 $\pm$ 727, 1925 $\pm$ 964, and 1431 $\pm$ 717 km/s.

The barometric model of EQ Peg predicts the range observable burst would reside is between 1.6 to 2.2 R$_{\star}$ under various conditions.  
This corresponds to a maximal expected areas of 0.17 and 0.41 A$_{\star}$.
A burst is expected to be observable over a range of $\sim$0.3 R$_{\star}$.  
The average velocity predicted by the four flares (1530 km/s) over this distance predicts drift rates between 5.03 and 7.04 MHz/s. 
Given the observation specifications, any burst would have appeared as essentially a vertical line in our dynamic spectrum.  

The predicted theoretical sensitivity for the observations was $\sim$33 mJy.  
The dynamic spectra RMS sensitivities were 115.0, 305.27, and 122.22 mJy.  
None of the dynamic spectra contained events that would be associated to a type II burst.  

Looking forward, telescopes like the proposed Next Generation Very Large Array would be able to probe stellar winds more deeply and further constrain mass loss rates due to winds and CMEs \citep{OstenCrosley2017}.
A shift in approach to sky surveys, throwing away specific stellar information prohibiting CME parameter constraints, such as the VLA Low Band Ionospheric and Transient Experiment \citep{Clarke2016} may be useful to first verify the existence of type II bursts before choosing specific targets to extrapolate constraints.  

\acknowledgements
MKC and RAO acknowledges funding support from NSF AST-1412525 for the project on which this paper is based.  
MKC and RAO thank NRAO director Anthony J. Beasley for the Directors Discretionary Time at the JVLA to complete this science.  


\end{document}